\documentstyle[aps,preprint]{revtex}

\begin{document}

\title{Energy Barriers for Flux Lines in 3 Dimensions}
\author{Barbara Drossel}
\address{Department of Physics, Massachusetts Institute of
Technology, Cambridge, Massachusetts 02139}
\date{\today}
\maketitle
\begin{abstract}
I determine the scaling behavior of the free energy barriers
encountered by a flux line in moving through  a
three-dimensional random potential. A combination of
numerical simulations and analytic arguments suggest that these
barriers scale with the length of the line in the same way
as the fluctuation in the free energy.

{\bf Keywords:} Flux lines, random media, energy barriers, glassy systems.
\end{abstract}

\pacs{PACS numbers: 74.60.Ge, 05.70.Ln, 05.40.+j}
\bigskip

Magnetic flux lines (FL) in high-$T_c$ superconductors are one of
the simplest examples of glassy systems \cite{glass1,glass2}.
In thermal equilibrium,
 a FL is pinned by defects (oxygen impurities, grain boundaries, etc)
in the superconductor which lower
its energy \cite{expreview}. This effect is limited by the line tension which
opposes the bending of the line. The resulting free energy
landscape for the FL is rather complicated and has many local
minima, i.e. metastable states \cite{BoseGlass}.
When an electric current flows
through the system, the FL feels a Lorentz force perpendicular to its
orientation and to the current direction.
As long as the current is not strong enough to overcome the pinning
forces, the line moves by thermally
activated jumps of line segments between metastable configurations
\cite{KimAnderson,FisherFisherHuse,JoffeVinokur}.
The length of these line segments is estimated by the condition
that the free energy barrier for a jump should be of the same
order as the gain in free energy due to that jump. These dynamics
are believed to be the reason for the nonlinear
voltage-current characteristics found in experiments \cite{expreview}.

Since energy barriers play such an important role in the dynamics
of FLs, it is essential to know their properties. The scale
of these barriers should grow with observation size $L$ like
a power law $L^\psi$. Usually, it is assumed that the energy
scale in the system is set by the fluctuations in free
energy which increase as
$L^\theta$, and that therefore $\psi = \theta$
\cite{HuseHenley,JoffeVinokur}. However,
it is also quite possible that the heights of the ridges in
the random energy landscape scale differently from those of the
valleys  that  they separate, with $\psi>\theta$.  Yet another
scenario is that transport occurs mainly along a percolating
channel of exceptionally low energy valleys with $\psi<\theta$.
A first attempt to clarify this situation was taken in \cite{MDK},
 where $\psi = \theta$ was established for a FL moving in 2 dimensions. This
demonstration relied on exact results for minimal energies
in $1+1$ dimensions, and on the fact that the endpoint of a FL in 2
dimensions has to  move through all points which
lie between its initial and final positions.
It is of great importance to discuss also a FL in 3
dimensions, which is more physically relevant. In contrast
to a two-dimensional system, a FL which moves in three
dimensions can avoid regions in space which are energetically unfavorable
for one of its segments, and one might therefore
speculate that $\psi<\theta$ in three dimensions. In this
paper, we first determine numerically a lower bound for the barrier
energy which scales in the same way as the energy fluctuations,
thus ruling out $\psi<\theta$. Further numerical results
 predict that an upper bound scales in the same way,
ruling out  $\psi>\theta$, and thus leading to $\psi = \theta$.

We describe the FL as a directed path in a random medium \cite{KardarRev}.
The path is discretized to lie on the bonds of a cubic lattice, starting
at the origin and  directed
along its (1,1,1) diagonal. Each segment of the line can proceed
in the positive direction along
one of the three axes, leading to a total of $3^t$
configurations  after $t$ steps, with endpoints lying in the plane
which is
spanned by the points $(t,0,0)$, $(0,t,0)$, and $(0,0,t)$.
 A given configuration of the FL is labelled
 by vectors $\left\{ \vec x(\tau)
\right\}$ for $\tau=0,1,\cdots,t$, giving the transverse coordinates
of the FL at each step. The
points  $\left\{ \vec x(\tau)\right\}$ lie on the
vertices of a triangular lattice. For a given value of $\tau$, they
lie on one of three alternating sublattices.

 To each bond on the cubic lattice is assigned a (quenched)
random energy equally distributed
between 0 and 1. The energy of each configuration is the sum of all
random bond energies on the line.
For each endpoint $(t,\vec x)$, there is a configuration of
minimal energy $E_{min}(\vec x |t)$ which can be obtained
numerically in a time of order $t^3$ by a transfer matrix algorithm
 \cite{KardarRev}.
The fluctuation in minimal
energy is known to scale as $t^\theta$ with $\theta \simeq 0.24$,
and the transverse fluctuation
of the coordinates of minimal paths is known to scale as $t^\zeta$,
with $\zeta \simeq 0.62$ \cite{BarabasiStanley}.
The endpoints of the minimal paths with the lowest energy lie
within a distance $\propto t^\zeta$ of the origin. Fig.~\ref{profile}
shows the minimal energies of paths of length $t = 288$ to
endpoints $\vec x$ with $|\vec x| < O(t^\zeta)$. The highest energy in
this region is represented in white, the smallest energy in black.
The minimal energies are correlated over
a distance of the order of $t^\zeta$. The distribution of minimal
energies is close to a Gaussian and is shown in Fig.~\ref{min}.
Similar to a 2-dimensional system \cite{HalpinHealy},
this distribution seems to have a third cumulant since it is not
completely symmetric.

We next examine the energy barrier that
has to be overcome when  the line is moved from an initial
minimal  energy configuration between $(0,\vec 0)$ and $(t,
\vec x_i)$ to a final one between  $(0,\vec 0)$ and
$(t,\vec x_f)$, with $|\vec x_{f,i}| \le t^\zeta$.
The only elementary move allowed is flipping a
kink along the line. Thus the point $(\tau, \vec x)$ can
be shifted to $(\tau, \vec x \pm \vec e_i)$, where $\pm \vec e_i$ are
the six vectors which connect a vertex in the triangular lattice
to its nearest neighbors within the same sublattice.
 Each route from the initial
to the  final configuration is
obtained by a sequence of such elementary moves.
For each sequence,  there
is an intermediate configuration of maximum energy, and a barrier
which
is the difference between this maximum and the initial energy.
In a  system at
temperature  $T$, the probability that the FL  chooses a sequence
which crosses a barrier of height $E_B$ is proportional to
$\exp(-E_B/T)$, multiplied by the number of such sequences.
We assume  that, as is the case for the equilibrium FL,
the ``entropic''
factor of the number of paths does not modify scaling behavior.
Thus at sufficiently low temperatures the FL chooses the optimal
sequence  which has to overcome the least energy, and the overall
barrier is the minimum of barrier energies of all sequences.

Since the number of elementary moves
scales roughly as the volume of a cone which contains
 the initial and final lines, the number of possible
sequences grows as $t^{x^2t}$. This exponential growth makes
it practically impossible to find the barrier by examining all
possible sequences,
hampering a systematic examination of barrier energies.
Rather than  finding the true barrier energy, we proceed by
placing lower and upper bounds on it.

A lower bound to the barrier energy is obtained in the following
way: While the line moves from its initial to its final
configuration, the transverse coordinates of its endpoint move
 between nearest-neighbor
positions on one of the above mentioned triangular sublattices.
When the endpoint is at a position $(\vec x)$,
the energy of the line is at least as large as the minimal energy
$E_{min}(\vec x |t)$. The maximum of
all these minimal energies along the trajectory of the endpoint,
minus the energy of the initial configuration, certainly bounds
the barrier energy from below. Since we do not know the actual
trajectory of the endpoint, we have to look for the trajectory
with the smallest maximal energy. Only in this case
we can be sure that we have indeed found a lower bound. This
situation is fundamentally different from a 2-dimensional system,
where there is only one possible trajectory for the endpoint.

Provided that the minimal energies $E_{min}(\vec x |t)$ are known,
this lower bound is determined
in polynomial time by using a transfer-matrix method:
We start by assigning to the initial point $\vec x_i$ a lower
bound energy 0, and to all other sites $\vec x$ on the same
sublattice an energy $t$
which is certainly larger than the lower bound resulting
 from the algorithm after many iterations. At each step the
energy of all sites $\vec x$ except the initial site is updated
according to the following rule: Look for the minimum of the energies
of the 6 neighbors $\vec x \pm \vec e_i$. If this is smaller than the
energy at $\vec x$, replace the energy at $\vec x$
by this minimum or by $E_{min}(\vec x|t) - E_{min}(\vec x_i|t)$,
whichever is larger. After a sufficiently large number of
iterations,
which is of the order of the size of the area of interest (which
scales as $t^{2\zeta}$), all possible trajectories to endpoints
within this area have been probed, and the energies do not
change any more. The energy at site $\vec x_f$ is then identified
as the lower bound. Fig.~\ref{lower} shows the lower bound to the
energy barrier for a line with the endpoint moving from the origin
to sites within a distance of the order of
$t^\zeta$, for different values of $t$ and averaged over 500
realizations of randomness. The distance $|\vec x_f - \vec x_i|$ has been
scaled by $t^{-\zeta}$, and the energy by $t^{-\theta}$. With
this scaling, all the curves should collapse, leading to
the following scaling behavior for the lower bound,
\begin{equation}
\left\langle {E_-} (t,|\vec x_f - \vec x_i|) \right\rangle   =
 t^{\theta} f_-(|\vec x_f - \vec x_i| / t^{\zeta}).\label{eq1}
\end{equation}
The function $f(y)$ is proportional to $y^{\theta/\zeta}$
for small $y$. For the simulated system sizes, however,
this asymptotic scaling cannot yet be clearly seen.
For $y > 1$, the scaling form in eq.~(\ref{eq1}) breaks down
since the minimal energy is then a function of the angle $(|\vec x| / t)$.
We conclude that the lower bound to the barrier scales in the same way
as the
fluctuations in minimal energy, and consequently the energy
barrier increases at least as $t^\theta$, leading to $\psi \ge
 \theta$.
The distribution $P(E_-)$ of the lower bound energy
for a given distance $|\vec x| \propto t^\zeta$ is
 shown in Fig.~\ref{distribution}.
It appears to be half-Gaussian with width $\propto t^\zeta$.

The result $\psi \ge  \theta$ is not surprising if one realizes
that an even simpler lower bound is given by $\max(E_{min}(\vec x_f|t)
- E_{min}(\vec x_i|t), 0)$, which evidently
scales as $t^\theta$ since the
distribution function of minimal energies decays exponentially
fast, i.e. has no power-law tails (see Fig.~\ref{min}).
To make sure that the scaling of the lower bound found above is not
dominated by the neighborhood of final configurations with particularly
high energies, I repeated the above simulations
 by allowing only endpoints with minimal
energies smaller than the initial energy.
This corresponds to the situation that the endpoint of the line does
not move to an arbitrary position but to a position which is
 energetically favorable.
The result
is shown in Fig.~\ref{lower} and has the scaling form
\begin{equation}
\left\langle {\tilde E_-} (t,|\vec x_f - \vec x_i|) \right\rangle   =
 t^{\theta} \tilde f_-(|\vec x_f - \vec x_i| / t^{\zeta}). \label{eq2}
\end{equation}
As in the previous case, the asymptotic scaling $\tilde f_-(y) \propto
y^{\theta/\zeta}$ for small $y$ cannot yet be clearly seen. The energy
distribution of the lower bound is again a half-Gaussian of width
$\propto t^{\zeta}$ and looks similar
to Fig.~\ref{distribution}.

The same scaling behavior is also found when instead of the
optimal trajectory for the endpoint the shortest trajectory
(a straight line) is chosen. In this case, the mean of the barrier
energy $E_0$ has the scaling form
\begin{equation}
\left\langle {\tilde E_0} (t,|\vec x_f - \vec x_i|) \right\rangle
 = t^{\theta} f_0( |\vec x_f - \vec x_i|/ t^{\zeta})\label{eq3}
\end{equation}
(see Fig.~\ref{lower}), again with a half-Gaussian
distribution of width $\propto t^\zeta$.
This, of course, does not represent a lower bound to the true
barrier, but it will be important for the determination of
an upper bound below, and is therefore included here.

The result $\tilde E_- \propto t^\theta$ (eq.~(\ref{eq2})) can be explained
from the exponential
tails of the distribution of minimal energies:
If we asume that the
endpoint of the line moves only in valleys of particularly low
energy, we can successively remove all sites with the largest
 minimal energy from the set of possible endpoints, until the connectivity
over the distance $t^\zeta$ breaks down.
The remaining endpoints form
 percolation clusters, and their density is given by the
corresponding percolation threshold (This is analogous to random resistor
networks describing the hopping resistivity for strongly localized electrons.
The resistance of the whole sample is governed by the critical resistor
that makes the network percolate \cite{randomresistor}). Since the occupied
sites are
correlated over the distances considered, the value for the
threshold is different from the site percolation threshold of 0.5 in
an infinite triangular lattice with no correlation between occupied
sites. But for the present purpose, it is sufficient to know that
this threshold is finite and that therefore a finite percentage
of all sites are below  threshold. Since the distribution of
minimal energies dacays fast, its tail cannot contain a finite
percentage of all sites. We conclude that the threshold is
within a distance of $t^\theta$ from the peak, and therefore that
 the energy fluctuation on the percolation cluster and
consequently the lower bound for the barrier are $\propto
t^\theta$.

We now proceed to construct an upper bound to the energy
barrier. To this purpose, we specify a sequence of
elementary moves which take the line from its initial to its
final configuration. Since we cannot be sure that this sequence
is the optimal one, we know only that the barrier associated
with this specific sequence is an upper bound to the true barrier.
The algorithm for the motion of the line is inspired by the
one presented in \cite{MDK} and is as follows: First, one choses
a sequence of endpoints connecting the initial to the
final endpoint which is as short as possible. Then, one draws
all the minimal paths leading to these endpoints. It is certainly
advantageous to keep the  intermediate paths as close to minimal
configurations as possible and therefore to require that the line
passes successively through all these intermediate minimal configurations.
Usually, minimal configurations $\{\vec x_1(\tau)\}$ and $\{\vec x_2(\tau)\}
$ with neighboring endpoints have
large parts in common and separate only during the last few steps.
They enclose a small loop with a size of the order of 1.
But sometimes, both paths already separate during the first few steps
and form a large loop of the lateral size of the order
of $t^\zeta$. We have to give a prescription for how the line
moves over a loop. If the two minimal paths have nowhere a distance
larger than 1 (measured in units of $|\vec e_i|$), we can choose a
sequence of elementary moves such that at most two bonds of the
line are not on one or the other minimal path, leading to a
barrier of order 1 between the two. If the distance is larger than
1, we proceed as follows: Let $\tau_0$ be the last point which
both lines have in common, i.e. $\vec x_1(\tau_0) = \vec x_2(\tau_0)$ and
$\vec x_1(\tau_0 + 1) \neq \vec x_2(\tau_0 + 1)$. We then consider the
midway points  ${(\tau_0 + (t - \tau_0)/2, \vec x)}$ which connect both
lines in the middle of the loop via the shortest possible
trajectory (if there are several possibilities, we choose one
at random).  For each of
these points, we find two minimal segments of length $(t - \tau_0)
/2$   connecting on one side
to $(\tau_0, \vec x_1(\tau_0)
)$ and on the other to either $(t, \vec x_1(t))$ or
$(t, \vec x_2(t))$.  There are usually several possibilities in
making these connections, but all of them lead to the same result.
The two segments form an almost minimal path of length $t - \tau_0
$,  constrained to go through
the point $(\tau_0 + (t - \tau_0)/2, \vec x)$.  We next move the line
$\{\vec x_1(\tau)\}$ with $\tau_0 \le \tau \le t$ stepwise through
this sequence of almost minimal paths. At each step we first
attempt to move the upper segment and then the lower one.
The prescription for moving these segments of length $(t - \tau_0)
/2$  is  exactly the same as for
paths of length $t$: If the distance between two consecutive
configurations is larger than
1 for some  $\tau$, we consider the points in the middle of the
loop formed by the two, and construct minimal paths of half the
loop  length  connecting them to the
initial and final loop points. Next we attempt to move segments
of the length of the loop by
repeatedly moving the upper and lower
line portion. In some cases, it
is necessary to proceed with this construction until the cutoff
scale  (1) is
reached. Thus, at each intermediate configuration the line is
composed of  segments of minimal paths of different length,
the smallest segment having length 1 in the worst case.

We now estimate the barrier energy resulting from the above
construction. In principle, this can be done by programming
the algorithm and determining the result numerically. Such a
program would be more complicated than in 2 dimensions, and it would not
be able to simulate large systems, since a big portion of the bond energies
on the three-dimensional lattice need to be stored. It is therefore
uncertain if the asymptotic scaling behavior
of the upper bound may be found this way.
 Instead, we resort to analytic considerations:
Since the line is always composed of segments of minimal paths,
whose scaling properties are known, we have enough information to
 give an upper bound to the
barrier energy. For small distances $|\vec x_f - \vec x_i|$, the initial
and final path usually differ only in the last few steps, and
therefore the upper bound to the barrier energy
(and also the barrier energy itself)
does not depend on $t$, but only on $|\vec x_f - \vec x_i|$.
We are interested in the barrier which has to be overcome when
the line moves over a distance of the order of $t^\zeta$. Let us
add successively the contributions to the upper bound which result from
the different steps in the algorithm: The first step consists
in finding a sequence of minimal paths with endpoints lying
on the shortest trajectory from $\vec x_i(t)$ to $\vec x_f(t)$. The energy
difference between the minimal path with highest energy and the
initial minimal path is $E_0(t,t^\zeta)\simeq t^\theta$, which
is the first contribution to the upper bound. When moving from one
minimal path to the next, the line has to overcome a loop which
in the worst case has the length $t$ and which might (again in the
worst case) occur in combination with the minimal path with highest
energy. We therefore have to add to the upper bound  the contribution
of a loop of length $t$ and  width $\propto t^\zeta$. This is obtained
as follows: Within the loop, the line
moves through a sequence of paths which are composed of two pieces
of minimal path of half the looplength. Both the upper and the
lower sequence cover an energy range $E_0(t/2,(t/2)^\zeta) \simeq
(1/2)^\theta E_0(t,t^\zeta)$. In the worst case,
both sequences have their maximum simultaneously, giving a
contribution $(1/2)^{\theta-1} E_0(t,t^\zeta)$ to the upper bound.
All further contributions can immediately be written down
because of the recursive definition of the algorithm: While the
upper and lower segments move through minimal configurations within
the loop, they in turn have to overcome loops which
in the worst case have the size $t/4$, and so on. The sum of all
these contributions, averaged over different realizations of randomness,
 is
\begin{eqnarray}
\langle E_c(t,t^\zeta)\rangle &=& \langle E_0(t,t^\zeta\rangle) + 2 \,
\langle E_0(t,t^\zeta)\rangle \,
((1/2)^\theta +
(1/4)^\theta + \ldots \nonumber\\
 &=& \langle E_0(t,t^\zeta)\rangle\, (2/(1-(1/2)^\theta - 1)
\simeq 12.0\, \langle E_0(t,t^\zeta)\rangle .
\end{eqnarray}
In principle, one has to add a constant which accounts for the
breakdown of the scaling form of the energy increase  for small
loops. But this constant is of the order of one and can be neglected
with respect to the terms which increase with  $t^\theta$.

There are several configurations of the path which are expected
to have the energy $E_c$. They pass through
loops of all sizes and are composed of one
minimal segment of length $t/2$, one of length $t/4$, etc; ending
with two smallest pieces of  length $1$. Each of these is a
candidate to be the barrier path in our algorithm. To obtain
the mean value for the maximum of their energies we need to know
their number and their distribution, especially in
the large-energy tail. The exact number of candidate barriers
is not known, but we can be sure that it increases no
faster than $t^{1+\zeta}$ which is  the order of the
total number of intermediate configurations of the path.
The energy distribution for the candidate barriers results from
the energy distribution of their segments: Each of the segments
of length $\tau_i$ has an approximately (half)-Gaussian
 energy distribution with a width of the
order of $\tau_i^\theta$.
Since the different segments are constructed through a
specific recursive procedure, they might not be independent.
If we want to be sure to establish an upper bound, we have to
 assume the worst case that they are completely
dependent, resulting in a variance
\begin{eqnarray}
{\rm var}\left( E_c(t,t^\zeta) \right)&\simeq&\log_2(t)
({\rm var}\left( E_0(t,t^\zeta)\right)+
2 {\rm
var}\left( E_0(t/2,(t,2)^\zeta)\right)+
\cdots)\nonumber\\
&\simeq& 8.7 \ln(t) \, {\rm var}\left( E_0(t,t^\zeta) \right)\propto \ln(t)\,
t^{2\theta}.
\end{eqnarray}

It can be checked easily that (for large $N$),
the maximum of $N$  independent
Gaussian variables of mean $a$ and variance $\sigma^2$,
is a gaussian  of mean
$a+\sigma\sqrt{2\ln N}$ and variance $\sigma^2/(2\ln N)$.
Since the  candidate barriers have large segments
in common, their energies are  not
independent. Assuming their independence, we overestimate
again the barrier energy, but of course
we still establish an upper bound to it.
Putting all contributions together and taking into account the behavior
for small $|\vec x_f - \vec x_i|$, we finally
 obtain the following estimate for
the upper bound in barrier energy:
\begin{eqnarray}\label{meanE+}
\left\langle E_+(|\vec x_f - \vec x_i|,t) \right\rangle&=& \left\langle
E_c(|\vec x_f - \vec x_i|,t)
\right\rangle
+\sqrt{2\ln N{\rm var} E_c(|\vec x_f - \vec x_i|,t)}\nonumber\\
&\simeq& \left(\ln t \right)t^{\theta} f_+(|\vec x_f - \vec x_i| /
t^{\zeta})\, .
\end{eqnarray}

To conclude, I have shown that the energy barrier encountered
by a FL moving in a $3d$ random medium has an upper and a lower
bound which both increase with $t^\theta$, except for
logarithmic corrections. From this follows that the
barrier itself increases with $t^\theta$, confirming
the hypothesis $\psi = \theta$.
Since the argument presented in this paper is mainly based on the
exponential tails of the minimal energy distribution, it can be expected
that the result $\psi = \theta$ holds also in higher dimensions,
provided that the distribution of minimal energies still has exponential tails.

{\bf Acknowledgements:}
I thank Mehran Kardar for very helpful discussions and comments
on the manuscript. This work was supported by the Deutsche
Forschungsgemeinschaft (DFG) under Contract No. Dr 300/1-1, and by the
NSF grant No. DMR-93-03667.

\begin{figure}
\caption{ Minimal energies of paths of length $t = 288$ to
endpoints $\vec x$ with $|\vec x| < O(t^\zeta)$. White: High energies; Black:
Low energies.
}
\label{profile}
\end{figure}

\begin{figure}
\caption{Probability distribution $P(E_{min})$
of minimal energies $E_{min}(\vec 0|144)$,
averaged over 50000 realizations of randomness. The solid line is a Gaussian
distribution.
}
\label{min}
\end{figure}

\begin{figure}
\caption{Scaling functions $f_-(y)$, $\tilde f_-(y)$, and $f_0(y)$
defined in eqs. (1) -- (3)
for $t = 72$ (solid), $t = 144$ (dotted), $t = 288$ (dashed),
$t = 576$ (long dashed), and $t = 1152$ (dot-dashed),
averaged over 500 realizations of randomness. The straight line has
the slope $\theta/\zeta = 0.39$.
}
\label{lower}
\end{figure}

\begin{figure}
\caption{
Probability distribution of $E_-$. The parameters
and symbols are the same as in Fig.3.
}
\label{distribution}
\end{figure}

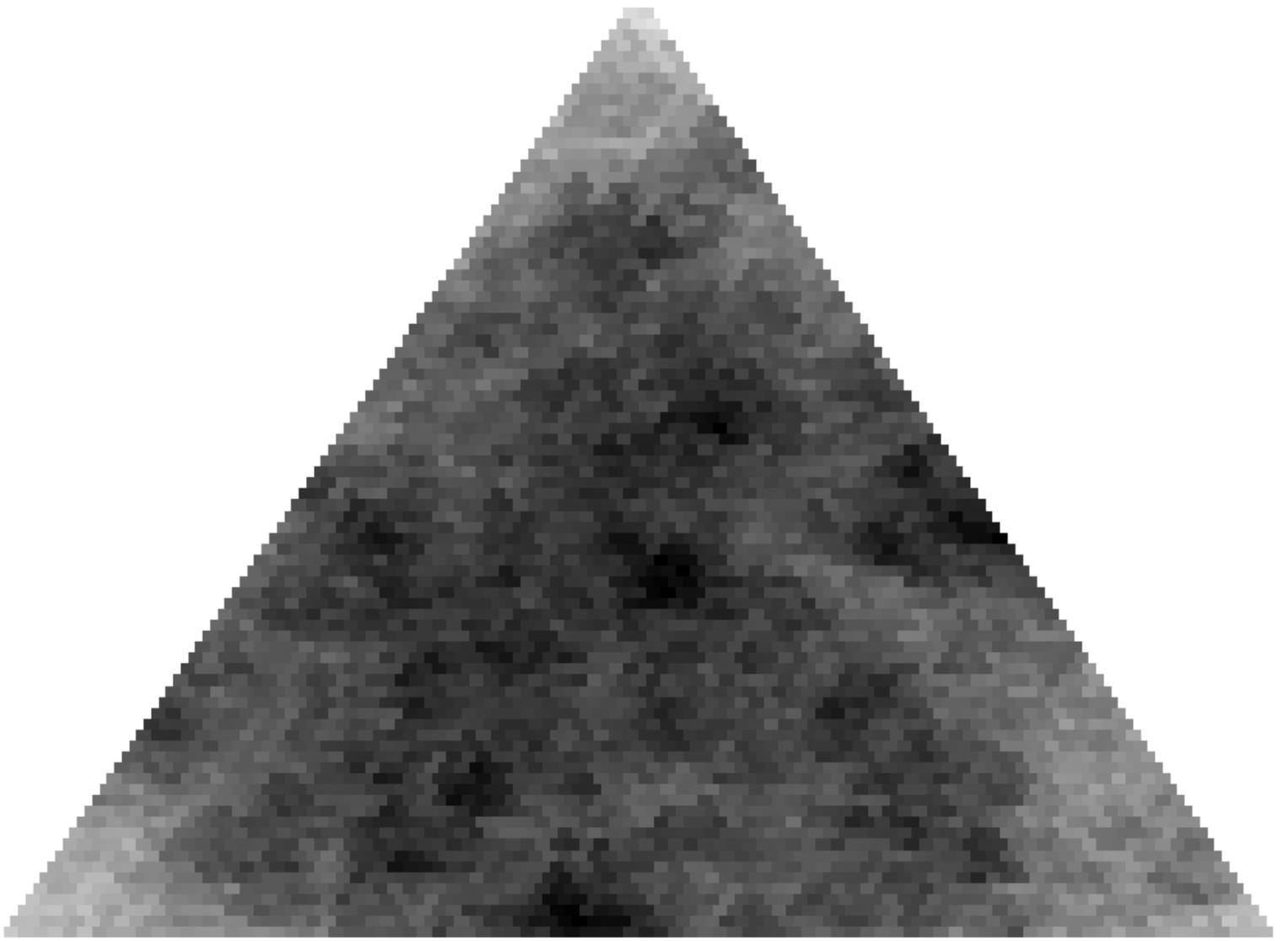
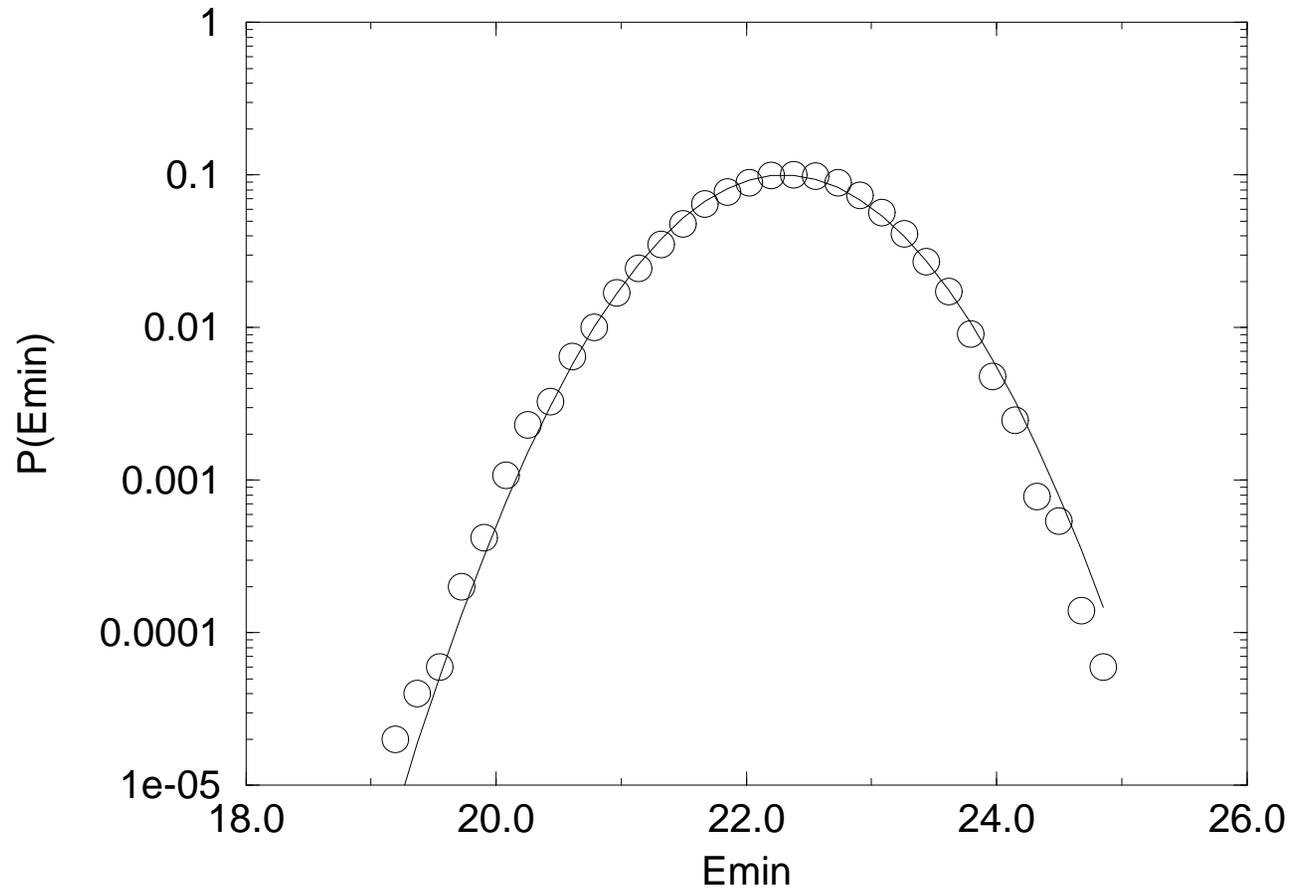
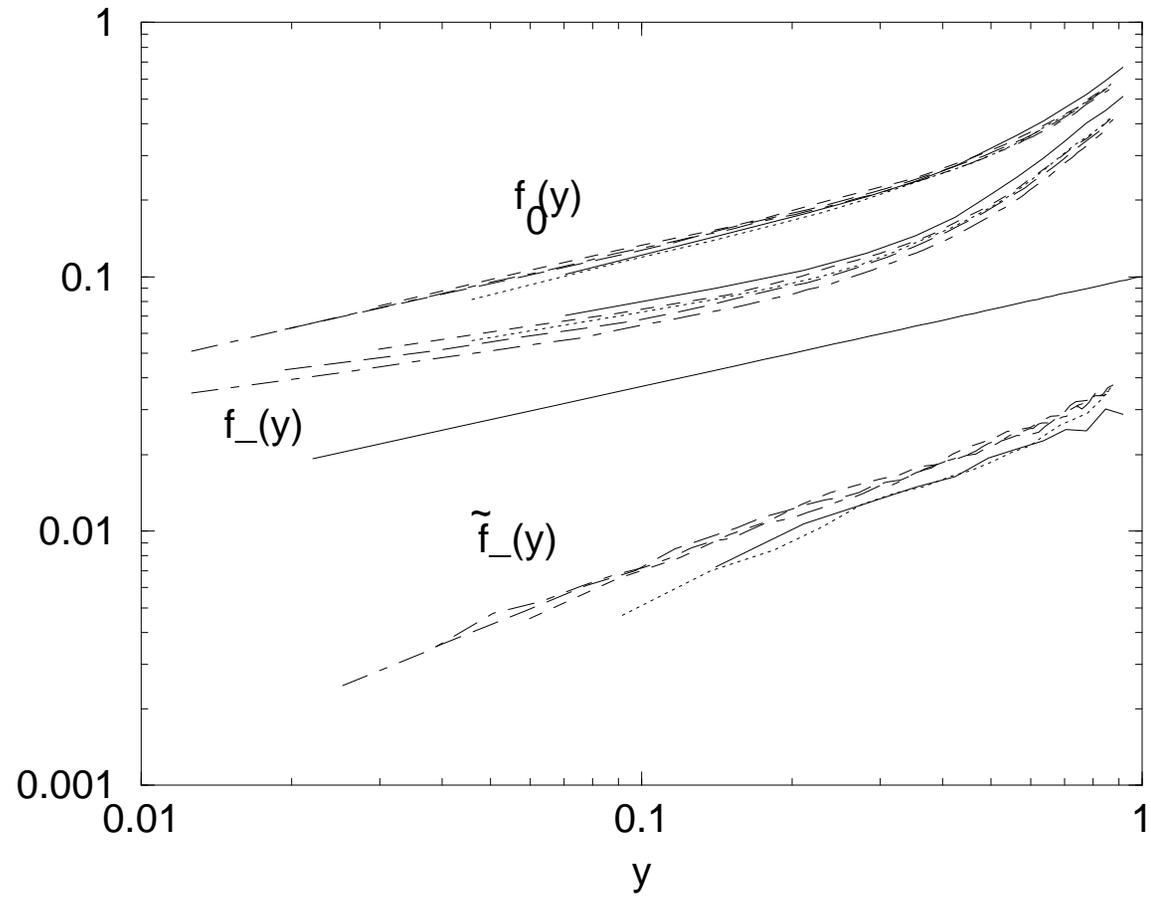
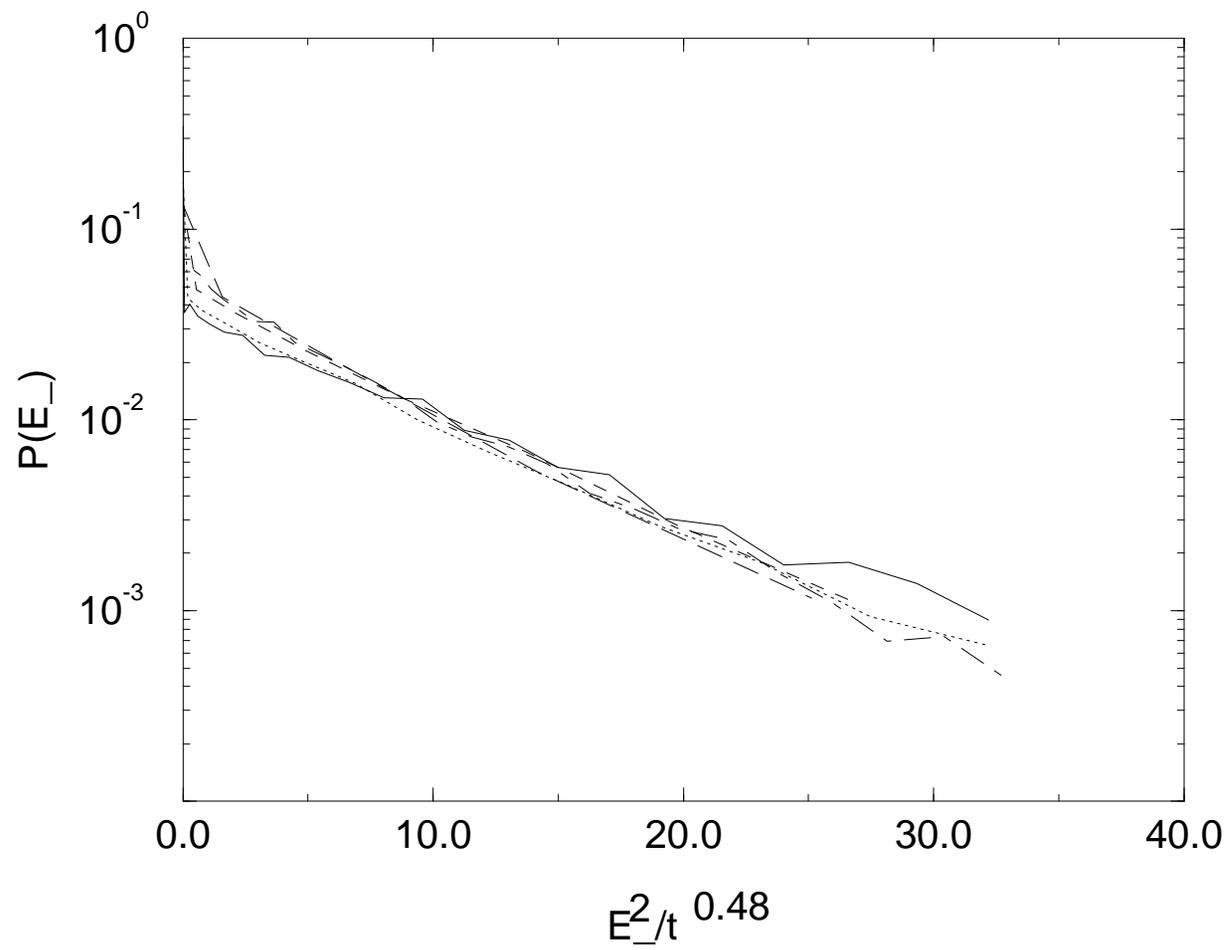
\end{document}